\documentclass[11pt]{article}

\usepackage[margin=1in]{geometry}
\usepackage{amsmath,amssymb}
\usepackage{booktabs}
\usepackage{array}
\usepackage{float}
\usepackage{threeparttable}
\usepackage{graphicx}
\usepackage{xcolor}
\usepackage[colorlinks=true, linkcolor=blue!60!black, citecolor=blue!60!black,
urlcolor=blue!60!black]{hyperref}
\usepackage[numbers]{natbib}
\usepackage{authblk}
\usepackage{enumitem}
\usepackage{caption}
\usepackage{subcaption}

\title{From Runnable to Verifiable: An Independent Reproducibility Study of
LLM/Agent-Driven Vulnerability Validation Artifacts}

\author[1]{Bo Chen}
\affil[1]{Institute of Computing Technology, Chinese Academy of Sciences}

\date{}

\begin{document}
\maketitle

\begin{abstract}
Security research artifacts---repositories, PoC exploits, and validation pipelines---are increasingly produced by LLM/agent-driven vulnerability workflows, yet the gap between \emph{publicly available}, \emph{runnable}, \emph{signal-producing}, and \emph{semantically confirmed} artifacts is poorly measured. We conduct a pre-registered reproducibility audit of this literature. A search covering 2023--2026 with dual screening yields a 104-paper consensus corpus, of which 59 papers (56.7\%) have a publicly reachable artifact. We execute an 18-paper sample at R0/R1 and all 102 cases of the anchor benchmark (arXiv:2509.24037), with patched-counterfactual verdicts on 30 signal-producing cases and matched-negative-control verdicts on 19. Three findings stand out. First, 58/102 (56.9\%) anchor cases contain a script-internal CVE identifier that diverges from the declared directory CVE. Second, only 10/18 (55.6\%) paper-level artifacts complete their declared workflow at R0, rising to 11/18 (61.1\%) after environment-only R1 repair. Third, artifact-embedded oracles prove unreliable: 20/30 patched-counterfactual audits still produce the claimed signal on the patched build, 7/19 matched negative controls still trigger on benign input, and the oracle confusion matrix has sensitivity 60\% and specificity 45\%. A trigger on the vulnerable build is not evidence of CVE-specific reproduction without a clean patched counterfactual. These are exploratory results from a pre-registered protocol, and our protocol---pre-registered post-conditions, R0/R1 repair ladder, G1--G3 semantic evidence levels, and patched-counterfactual oracles---is a reusable template for the security reproducibility community.
\end{abstract}

\section{Introduction}
\label{sec:intro}

Automated vulnerability validation by LLMs and agentic systems has advanced rapidly: agents that reproduce known CVEs, generate PoCs, repair vulnerable code, and validate exploits. However, the artifacts accompanying these papers vary widely in maturity, and we distinguish four levels between mere availability and verified reproduction. At the first level, \emph{available}, a primary artifact---repository or archive---exists and is publicly reachable. At the second, \emph{runnable}, the artifact completes its intended workflow on a clean snapshot (R0) or after minimal environment-compatible repair (R1). At the third, \emph{signal-producing}, execution yields candidate signals such as crashes, status changes, or success markers. At the fourth, \emph{semantically confirmed}, the signal matches a CVE-specific post-condition with a matched negative control and a patched counterfactual. Prior security-reproducibility studies measure general availability and runnability; none, to our knowledge, operationalize the \emph{semantic} confirmation gap for LLM/agent validation artifacts with per-case oracle audits.

Our contributions are threefold:
\begin{itemize}
\item A pre-registered, reproducible measurement protocol---CVE-specific post-conditions, an R0/R1 repair ladder, G1--G3 semantic evidence levels, and patched-counterfactual oracles---applied to a 104-paper consensus corpus: an 18-paper paper-level sample and a 102-case case-level sample.
\item Three findings from executing the full 102-case anchor corpus plus an 18-paper sample: script-internal CVE identifiers diverge from declared targets in 58/102 (56.9\%) cases; only 11/18 (61.1\%) paper-level artifacts complete their workflow after R0/R1; and artifact-embedded oracles are systematically unreliable---20/30 audits dirty, 7/19 controls false-positive, oracle-matrix sensitivity 60\% and specificity 45\%.
\item Open evidence bundles---frozen protocols, screening decisions, claim sheets, and per-case execution logs---as a reusable template for the security reproducibility community.
\end{itemize}

\section{Study Design and Pre-registration}
\label{sec:design}

\subsection{Research Questions}
We address three confirmatory research questions: RQ1 at the paper level, RQ2 and RQ3 at the case level. Artifact availability (Section~\ref{sec:availability}) is reported as a census of the consensus corpus rather than a separate research question, and the calibration results of Section~\ref{sec:calibration} are explicitly exploratory.
\begin{description}[leftmargin=2.2em]
  \item[RQ1] Among eligible papers with accessible artifacts, what fraction complete their intended workflow under R0 and R1, and what are the dominant failure modes and repair burdens (time, diffs, severity)?
  \item[RQ2] How large is the gap between papers' claimed per-case success and independently confirmed E1 evidence?
  \item[RQ3] How reliable are artifact-embedded oracles against vulnerable/patched ground truth (TP/FP/FN/TN)?
\end{description}

\subsection{Execution Modes and Evidence Levels}
Every artifact is executed in one of two modes: R0 on a clean snapshot with zero modification, or R1 after environment-compatible repair only (exploit, oracle, and post-condition untouched); semantic rewrites of the attack are reserved for case studies and were not needed here. 

Evidence levels grade what a produced signal establishes about the claimed CVE and operationalize the four conceptual levels of Section~\ref{sec:intro}. The G-levels form an increasing-evidence ladder---each condition, when met, strengthens the case that the signal is CVE-specific---and each answers a distinct question:
\begin{itemize}
\item \textbf{G1, candidate signal}: crash, marker, or status change; the signal-producing level.
\item \textbf{G2, post-condition match}: the signal matches the CVE behavior.
\item \textbf{G3a, negative control}: the signal is input-independent---absent on benign input.
\item \textbf{G3b, patched counterfactual}: the signal is version-independent---absent on the patched build (a parallel dimension to G3a rather than a further step).
\end{itemize}
The E-levels are verdicts derived from the G-conditions:
\begin{itemize}
\item \textbf{E1} = G2+G3a+G3b: strict confirmation.
\item \textbf{E2} = G2+G3a: confirmation without patch.
\item \textbf{E3}: incomplete evidence.
\end{itemize}
A case can record partial evidence (e.g., G2 met with G3b pending) without collapsing into a failure verdict; the same ladder applies to artifact-embedded oracles and to our independent assessment.

\subsection{Statistical Analysis Plan}
This plan was pre-registered before any confirmatory execution, to prevent choosing tests after seeing the data. Results are reported in Section~\ref{sec:calibration}. Analyses that need multi-paper pool are deferred to future work (Section~\ref{sec:conclusion}).
\begin{itemize}
\item \textbf{Primary unit}: paper. Per-paper rates aggregated as median/IQR/range.
\item \textbf{Case-level inference}: cluster bootstrap $\geq$2000 resamples with paper as cluster.
\item \textbf{Claim gap}: paired claimed vs E1 confirmation; Wilcoxon signed-rank + Hodges--Lehmann; descriptive-only reporting when fewer than 10 papers.
\item \textbf{Oracle audit}: confusion matrix with Wilson 95\% CI; E1 primary, E1+E2 sensitivity.
\item \textbf{Failure taxonomy}: mutually exclusive primary cause; multi-label secondary; prevalence + co-occurrence.
\item \textbf{Unknowns}: best/worst-case bounds, no single-value imputation.
\end{itemize}

\section{Corpus Construction and Artifact Availability}
\label{sec:search}

\subsection{Search and Screening}
The search protocol is frozen (\texttt{protocol/search\_protocol\_v1.json}). It covers the window 2023-01-01 to 2026-08-08, using the core query \texttt{(LLM OR agent) AND (vulnerability OR CVE OR exploit OR PoC) AND (reproduce OR validate OR verification OR exploit generation)}. Automated queries were executed against arXiv, DBLP, and OpenAlex (the latter two index ACM, IEEE, USENIX, and NDSS publications indirectly, contributing most non-arXiv candidates).

\begin{figure}[H]
\centering
\includegraphics[width=0.8\linewidth]{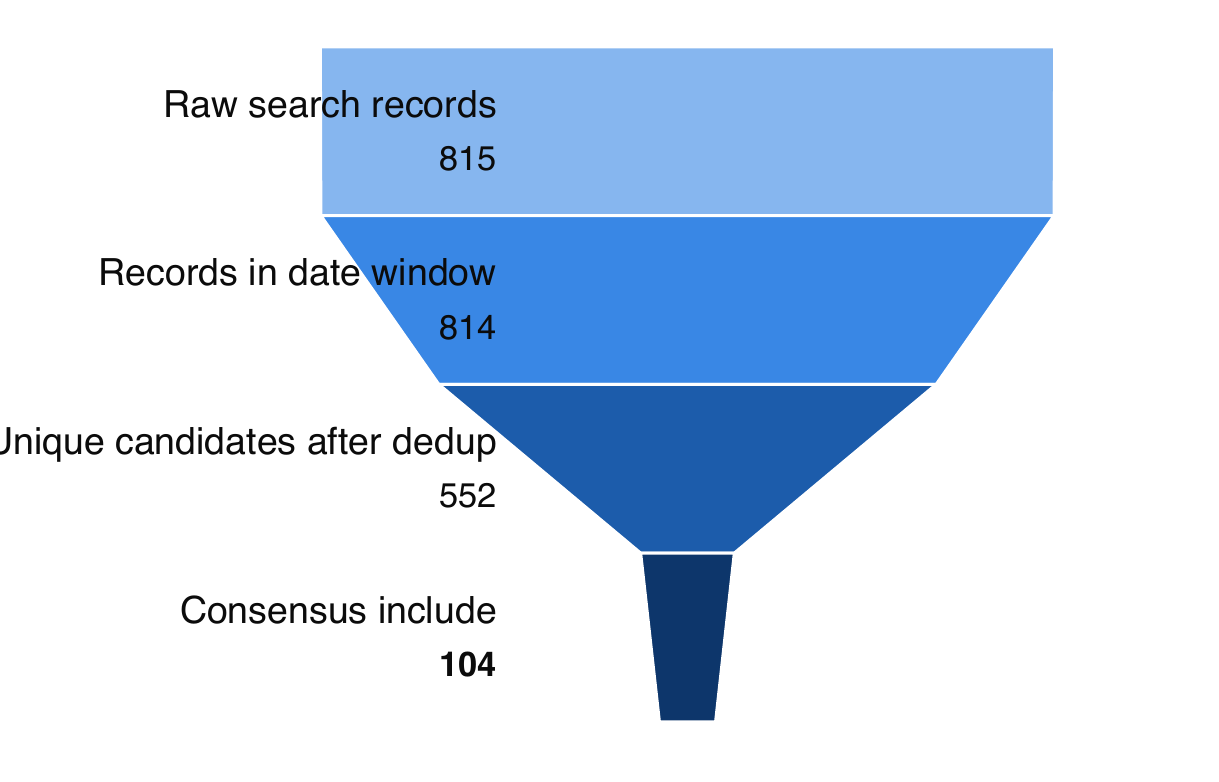}
\caption{Screening funnel: raw search records to the 104-paper consensus corpus (frozen 2026-08-08).}
\label{fig:funnel}
\end{figure}

The queries returned 815 raw records, of which 814 fell in the date window and 552 remained after deduplication. Applying five mutually exclusive exclusion rules (Table~\ref{tab:exclusions}) to the 552 candidates---no real vulnerability (249), publication type mismatch (94), no LLM/agent (59), no execution validation (40), and no extractable claim (6)---yielded the 104-paper consensus corpus (Figure~\ref{fig:funnel}).

\begin{table}[H]
\centering
\caption{Exclusion rules: mutually exclusive primary reasons.}
\label{tab:exclusions}
\footnotesize
\begin{tabular}{@{}lp{4.6cm}rl@{}}
\toprule
Rule & Definition & $N$ & Examples \\
\midrule
No real vulnerability & object is not a real CVE: capability evaluations, theoretical analyses, cross-domain LLM studies & 249 & exploit-capability evaluations \\
Publication type mismatch & non-research publication: survey, report, opinion, position & 94 & overviews; technical reports \\
No LLM/agent & core method uses no LLM/agent: reinforcement learning, statistical models, datasets & 59 & RL pentesting; neural repair \\
No execution validation & no execution validation of the claimed workflow: framework, proposal, static analysis & 40 & detection frameworks \\
No extractable claim & no extractable per-case success claim & 6 & toolkits \\
\bottomrule
\end{tabular}
\end{table}

\subsection{Artifact Availability}
\label{sec:availability}

\begin{figure}[H]
\centering
\includegraphics[width=0.92\linewidth]{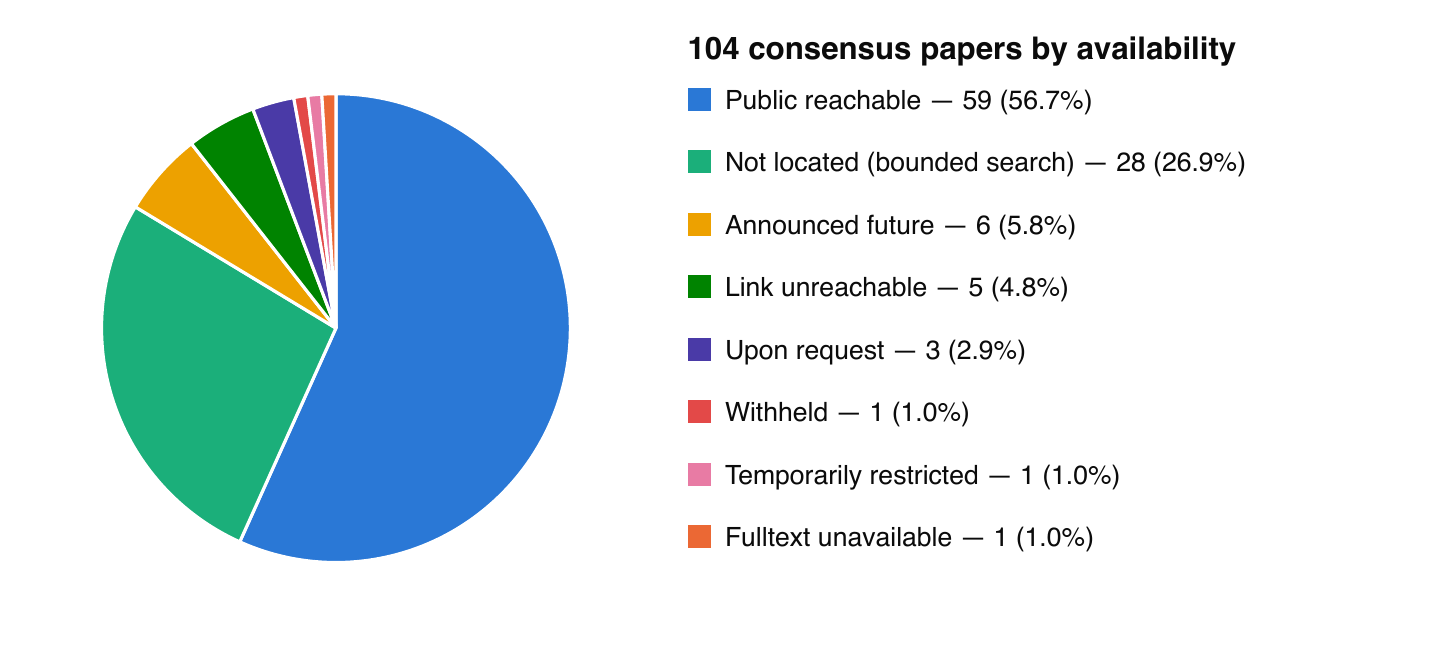}
\caption{Artifact availability on the consensus set ($n=104$, audited 2026-08-08).}
\label{fig:availability}
\end{figure}

The availability audit of the 104 consensus papers produced the classes of Figure~\ref{fig:availability}: 59/104 (56.7\%) had a publicly reachable primary artifact; the audit procedure is detailed below.

Two artifacts hosted on 4open.science (a platform for anonymous-review repositories, where authors upload artifacts under a temporary anonymous link during blind review)---Excalibur and Attention\_Distance---were reachable at initial audit but returned \texttt{repository expired} at acquisition time; we reclassified them as \texttt{link unreachable}. This illustrates the temporality of anonymous-review artifacts: such links lapse after the review cycle, so an artifact can be reachable when audited yet gone by the time a reader acquires it. Acquisition also surfaced a second form of availability temporality: one anonymous repository (PAGENT) was found to be a copy of the FaultLine benchmark repository~\citep{faultline}. Both observations feed the availability census and RQ1's failure taxonomy---artifacts that are reachable but not durably accessible.

The availability audit proceeds in three steps: primary artifact attribution from the paper fulltext and publisher pages; remote reachability checks via live probes (GitHub API, Zenodo/Figshare APIs, HTTP status); and a bounded negative search for not-located papers (paper PDF, arXiv record, exact-title web search, GitHub repo search). Acquisition freezes commit/archive hashes; reachable $\neq$ runnable.

\section{Sample Design}
\label{sec:samples}

The confirmatory design pairs two complementary samples, both frozen before any confirmatory execution: a paper-level sample that measures R0/R1 outcomes, failure modes, and repair burdens (RQ1), and a case-level sample that measures claim-to-evidence gaps and oracle reliability (RQ2, RQ3). All 104 consensus papers contribute to the availability statistics of Section~\ref{sec:availability}. The two samples select execution subsets from the 59 reachable papers.

\subsection{Paper-Level Sample}
\label{sec:paperlevel}

The paper-level sample comprises 18 papers drawn from the 59 publicly reachable papers, stratified by year (2023:1, 2024:6, 2025:6, 2026:5) and spanning repair, penetration testing, benchmarks, and smart contracts, drawn with deterministic selection seed 20260808. All 18 artifacts were acquired and commit-frozen. Their R0/R1 execution results are reported in Section~\ref{sec:paperlevel-results}; the full sample is listed in Appendix~\ref{app:paperlist}.

\subsection{Case-Level Sample}
\label{sec:caselevel}

The case-level sample audits individual cases: every executed case is scored against a pre-registered, CVE-specific post-condition. The paper pool comprises 42 papers---the subset of the consensus set whose primary artifacts were public reachable at audit time and were acquired and commit-frozen. The pool is the sampling frame for case-level execution and the cluster universe for the confirmatory statistics (paper as cluster). To date, one of its 42 papers---the anchor---has been executed at case level (its 102-case artifact is the current corpus), and cross-paper audits, initiated on the CVE-Bench corpus (also in the pool), extend into the remaining papers.

\subsection{Anchor Corpus and a Pre-registration Finding}
\label{sec:anchor}
The primary execution corpus is the 102-case anchor benchmark~\citep{anchor2509} (arXiv:2509.24037), one of the 42 papers in the case-level pool (Section~\ref{sec:caselevel}). Its artifact is unusually well suited to case-level auditing: 102 case directories, each shipping a \texttt{build\_and\_run.sh} and a README declaring PASS/FAIL (71 claimed PASS, 31 claimed FAIL; languages Python 41, JavaScript 16, Rust 15, C 12, Go 9, Shell 4, Java 2, C++ 2, Erlang 1), with per-case Docker setups. Its declared per-case claims make it the natural target for the calibration pilot. We pre-registered CVE-specific post-conditions for all 102 cases from NVD semantics \emph{before} any execution. A static consistency audit found that \textbf{58/102 (56.9\%)} cases contain a script-internal \texttt{CVE\_ID} variable that differs from the directory's declared CVE; per-case inspection shows this is not merely cosmetic---e.g., the directory declared CVE-2023-25668 runs a QuantizeAndDequantizeV2 trigger (CVE-2023-25676)~\citep{nvd202325676}, and the directory declared CVE-2021-28878 tests \texttt{Vec::from\_iter} (CVE-2021-31162)~\citep{nvd202131162}. This means the claimed reproduction target and the actually tested target may diverge, directly relevant to RQ2. 

From this corpus, a 20-case sample (5 crash PASS, 5 non-crash PASS, 5 code-generation FAIL, 5 environment FAIL) is frozen with pre-registered post-conditions derived from NVD descriptions. Stratification deliberately balances claimed PASS and FAIL (10/10) rather than mirroring the corpus-wide split (71 PASS / 31 FAIL), so that the protocol pilot exercises both claimed outcome classes; confirmatory case sampling will respect the corpus proportions. Notably, the anchor paper's own README marks the ``Postcond.'' column as absent for nearly all PASS cases, i.e., the paper never verified CVE-specific post-conditions---the exact gap this study measures. For the 58/102 cases with script-internal CVE divergence above, the claimed reproduction target and the actual test target must be reconciled before any R0 verdict.

\section{Results}
\label{sec:calibration}
This section reports the completed execution levels: the paper-level R0/R1 sample of 18 papers (Section~\ref{sec:paperlevel-results}) and the case-level anchor corpus of 102 cases. All results are \emph{exploratory}: the pre-registered confirmatory statistics require the multi-paper case-level pool that the cross-paper phase is building (Section~\ref{sec:caselevel}). Because the sub-analyses run on different sample sizes, Table~\ref{tab:execution} gives the execution ledger---every denominator used in this section and its relationship to the others.

\begin{table}[h]
\centering
\caption{Execution ledger: denominators for every sub-analysis.}
\label{tab:execution}
\footnotesize
\begin{tabular}{@{}lrll@{}}
\toprule
Stage & $N$ & Denominator note & Status \\
\midrule
Paper-level R0/R1 & 18 & of 18 papers & complete \\
Anchor corpus total & 102 & claim sheet (frozen) & --- \\
Acquired & 102 & all case dirs in repo & acquired \\
Calibration sample (frozen) & 20 & stratified 5$\times$4 & frozen \\
Calibration executed & 20 & all 20 & executed \\
New cases beyond calibration & 82 & 102 $-$ 20 & executed \\
Full-corpus runs & 87 & 82 new + 5 reruns & executed \\
Total cases executed & 102 & 20 calibration + 82 new & executed \\
G1 signals & 34 & of 87 runs (39.1\%) & detected \\
G2 verdicts & 23 met / 11 not & of 34 G1 & complete \\
G3b audited & 33 & 29/34 G1 + 4 cal.-only; 30 verdicts, 3 unbuild. & complete \\
G3b unaudited (of 34 G1) & 5 & 5 G1 cases not patched-audited & pending \\
G3b clean / dirty & 10 / 20 & of 30 verdicts & complete \\
G3a executed & 19 & of 34 G1 (9 pass / 7 FP / 2 N/A / 1 unassessable) & in progress \\
E1 confirmed & 2 & G2 met $\cap$ G3b clean $\cap$ G3a pass & derived \\
\bottomrule
\end{tabular}
\end{table}

\subsection{Paper-level R0/R1 execution}
\label{sec:paperlevel-results}
Each artifact was executed at R0 (clean snapshot) and, where the declared entry could not complete, at R1 (environment-only repair: missing dependencies installed from the artifact's declared manifest; exploit, oracle, and post-condition untouched). 10/18 artifacts (55.6\%) completed their declared workflow at R0; one further artifact (ContractTinker) completed after R1 repair; the remaining 7 failed at both levels. The dominant failure modes were: declared dependency manifests absent from the artifact root (5 artifacts), pip-incompatible manifests (uv-based projects; 2 artifacts), and missing runtime dependencies that were installed but did not suffice because the workflow additionally depends on external services or datasets (4 artifacts). R1 repairs were environment-only and small: installing declared dependencies took 7--132 seconds (median 42 s) with one to three pip commands, severity rated low to medium; no repair touched exploit, oracle, or post-condition code.

\subsection{Calibration R0 execution}
\label{sec:calib-r0}
The 20-case calibration sample frozen in Section~\ref{sec:anchor} was executed as a protocol pilot: all 20 cases ran at R0 (clean snapshot, Docker Desktop LinuxKit VM, resource-limited, privileged-disabled, network-policy per case). Table~\ref{tab:calibration} shows the pilot at R0; only 3/10 claimed-PASS cases produced a trigger signal on a clean snapshot, and most failures trace to EOL base-image apt sources, PoC compile errors, or artifact defects.
\begin{table}[h]
\centering
\caption{Claimed outcome vs.\ R0 trigger signal (20-case calibration).}
\label{tab:calibration}
\begin{tabular}{@{}lccc@{}}
\toprule
Claimed & $N$ & R0 trigger & R0 failed/no-trigger \\
\midrule
PASS & 10 & 3 & 7 \\
FAIL & 9 & 3 & 6 \\
env-fail (DNS) & 1 & 0 & 1 \\
\bottomrule
\end{tabular}

\smallskip
\noindent\footnotesize Rows are by claimed outcome; ``env-fail (DNS)'' is
CVE-2025-32433. Counts are R0-only; three further cases produced a signal
only after R1 repair (Section~\ref{sec:calib-r0}).
\normalsize
\end{table}

Claim-target mismatch also materializes at execution: several claimed-PASS cases whose scripts contain a different inner CVE identifier (Section~\ref{sec:anchor}) executed a different target than the directory claims. This divergence is distinct from ``wrong patch'' or EOL failures: it undermines the claimed reproduction before execution even begins, and it is invisible to availability audits and to automated oracles that only check for a trigger string.

R1 repair, where needed, was minimal and environment-only: the three R1-recovered cases (CVE-2021-31162, CVE-2022-35986, CVE-2023-25669) were fixed by redirecting EOL base-image apt sources to \texttt{archive.debian.org} and adjusting bash-4 incompatibilities in the build script---one-to three-line changes to the Dockerfile or build script, with exploit, oracle, and post-condition untouched. Per-repair wall-clock times were not recorded in the pilot; the paper-level sample quantifies repair burdens instead (7--132 s, one to three pip commands; Section~\ref{sec:paperlevel}).

CVE-2021-31162's artifact oracle reported ``TRIGGERED'' on both the vulnerable (rust 1.51) and patched (rust 1.52) builds ---an oracle false positive driven by a designed-in panic being mistaken for the vulnerability signal. CVE-2021-44228's payload executed (curl exit 0, ``Hello, world!'') but the oracle required a marker file that was never created, so the run was reported as not triggered despite a live JNDI RCE signal~\citep{nvd202144228}. Across the 20 executed calibration cases, 2 artifact oracles produced false positives (signal on vulnerable \emph{and} patched builds) and 3 produced false negatives (real signal not recognized)---oracle unreliability in 5/20 (25\%).

\subsection{Full-corpus R0 execution}
Beyond the calibration sample, the 82 anchor cases not in the calibration sample were executed at R0 under the same environment and safety gates (five calibration cases were additionally re-run for pipeline verification, 87 executed runs in total, covering all 102 cases together with the calibration). Signal detection used the frozen trigger definition: a run counts as G1 if it completes (exit 0) and its log contains a vulnerability-signal marker from a fixed family---script-internal oracle verdicts (``VULNERABILITY TRIGGERED''), crash markers (Segmentation fault, SIGSEGV, SIGFPE, Floating point exception), memory-safety reports (double-free, use-after-free, AddressSanitizer/UndefinedBehaviorSanitizer reports, heap- or stack-buffer-overflow), and runtime failures (panicked at, assertion failed, runtime error). Generic Python tracebacks or exit-code text without such a marker are not counted. 34 of the executed runs (39.1\%) produced a candidate signal (G1). G2 adjudication is complete: 23 of 34 signals satisfy the pre-registered post-condition (15 script-internal oracle verdicts whose trigger logic matches the CVE behavior, 5 keyword-family matches, 3 crash-signal correspondences), and 11 do not---including CVE-2025-0182, whose script-internal oracle prints its trigger unconditionally without verifying the claimed denial-of-service behavior. Of the 23 satisfying cases, one (CVE-2024-27936) is negated by its dirty patched counterfactual (Section~\ref{sec:calibration}). The per-case runs, trigger-logic excerpts, and adjudication records are archived in the evidence bundles.

\subsection{Patched-counterfactual (G3b) audit}
\label{sec:g3b}
G3b disproves a ``textbook'' success: CVE-2020-1967 produced a clean segmentation-fault signal (R0, exit 0) and matched the pre-registered post-condition (G2). Yet the patched counterfactual (OpenSSL 1.1.1g, the official fix) also segfaulted: the PoC calls \texttt{SSL\_check\_chain(None)} directly---a NULL-argument crash that is undefined behavior in \emph{any} OpenSSL version, not the real CVE-2020-1967 trigger (malformed TLS~1.3 \texttt{signature\_algorithms\_cert} extension)~\citep{nvd20201967}. The PoC's own comment says ``simulate vulnerability''. Thus the signal is not CVE-specific: \textbf{available $\to$ runnable $\to$ signal-producing did not imply semantically confirmed}.

The G3b-executed set within the calibration was a feasibility subset of five cases; it is not a random sample, so the 4/5 rate below is not an estimate of a population proportion. Of the 5 calibration cases executed against their patched build, 4 (80\%) \emph{still produced the claimed signal} on the patched version (CVE-2020-1967, CVE-2021-31162, CVE-2022-22816, CVE-2023-0217): three PoCs manually force corruption (NULL argument, ``pkey->ameth = NULL''~\citep{nvd20230217}, caught-exception-as-trigger~\citep{nvd202222816}) and crash in \emph{any} version; one oracle treats a designed-in panic as the signal. Only CVE-2025-30223 (Beego XSS) passed G3b: the patched 2.3.6 HTML-encodes the payload while 2.3.5 reflects it raw~\citep{nvd202530223}, yielding the single E1-level confirmation in the calibration set. \textbf{A trigger signal on the vulnerable build is not evidence of CVE-specific reproduction unless the patched counterfactual is clean.}

Of the 34 G1 signals from the full-corpus runs, 29 were patched-audited and 5 were not. Four additional calibration-only cases (outside the 34 full-corpus G1 signals) were also patched-audited, bringing the G3b-executed set to 33 cases. Patched builds were constructed per case (image-tag replacement for language runtimes, pinned package/crate version replacement for library-level fixes, and apt upgrade injection for system-library fixes, with large TensorFlow base images pre-pulled to make the builds feasible). Verdicts were reached on 30 cases: \textbf{10 clean} (signal absent on the patched build) and \textbf{20 dirty} (signal persists)---including 5 TensorFlow cases, 2 Go cases whose base images already contain the fix yet the signal persists (itself evidence of claim-target or PoC inconsistency), and apt-level cases whose EOL sources carry no fix. Three cases are recorded as \emph{unbuildable} after repeated attempts: a Rust crate with no patched release, an absl/protobuf link conflict in the build chain, and an apt proxy network failure. The overall G3b ledger---30 verdicts (10 clean / 20 dirty) plus 3 unbuildable---stands in sharp contrast to the artifacts' own PASS/FAIL claims: two-thirds of signal-producing cases do not survive their own patched counterfactual. Matched negative controls (G3a) were executed on 19 of the 34 signal-producing cases: 9 passed, 7 revealed oracle false positives, 2 were not applicable (no input differentiation in the vulnerability semantics), and 1 was unassessable (build-chain link error independent of the control input); the remaining cases require per-case PoC analysis.

\subsection{Outcome table}
Table~\ref{tab:g1outcomes} lists the adjudicated outcome for every G1 signal case. G2 marks whether the signal satisfies its pre-registered post-condition; G3b marks the patched-counterfactual verdict; G3a marks the matched-negative-control verdict; E1 marks the two cases with strict confirmation (G2 met $\cap$ G3b clean $\cap$ G3a pass). Four of the 20 dirty G3b verdicts (CVE-2020-1967, CVE-2021-31162, CVE-2022-22816, CVE-2023-0217) come from the calibration pilot and are not among the 34 full-corpus G1 signals, so they do not appear in the table.

\begin{table}[H]\centering\caption{Outcomes for all 34 G1 signal cases (G2 adjudication, G3b patched counterfactual, G3a matched control).}\label{tab:g1outcomes}\footnotesize
\begin{tabular}{@{}lllll@{}}\toprule
Case & G2 & G3b & G3a & E1 \\ \midrule
CVE-2021-29922 & $\checkmark$ & clean & FP &  \\
CVE-2021-42576 & $\checkmark$ & clean & pass & $\checkmark$ \\
CVE-2022-22817 & $\checkmark$ & clean & FP &  \\
CVE-2022-25219 & $\checkmark$ & dirty & pass &  \\
CVE-2022-25235 & $\checkmark$ & dirty & pass &  \\
CVE-2022-3358 & $\checkmark$ & dirty & --- &  \\
CVE-2022-35934 & $\times$ & clean & --- &  \\
CVE-2022-41885 & $\times$ & dirty & --- &  \\
CVE-2023-25663 & $\times$ & dirty & --- &  \\
CVE-2023-25664 & $\times$ & dirty & --- &  \\
CVE-2023-25665 & $\checkmark$ & dirty & --- &  \\
CVE-2023-25667 & $\checkmark$ & dirty & --- &  \\
CVE-2023-25672 & $\times$ & clean & --- &  \\
CVE-2023-25675 & $\checkmark$ & dirty & --- &  \\
CVE-2023-25801 & $\times$ & clean & --- &  \\
CVE-2023-28487 & $\checkmark$ & dirty & FP &  \\
CVE-2023-39325 & $\times$ & dirty & FP &  \\
CVE-2023-49210 & $\checkmark$ & --- & --- &  \\
CVE-2023-49292 & $\checkmark$ & --- & pass &  \\
CVE-2024-1597 & $\checkmark$ & --- & pass &  \\
CVE-2024-2410 & $\checkmark$ & unbuild. & unass. &  \\
CVE-2024-24790 & $\checkmark$ & dirty & FP &  \\
CVE-2024-27931 & $\times$ & --- & --- &  \\
CVE-2024-27934 & $\times$ & clean & pass &  \\
CVE-2024-27936 & $\checkmark$ & dirty & FP &  \\
CVE-2024-4340 & $\times$ & clean & pass &  \\
CVE-2024-5991 & $\checkmark$ & dirty & --- &  \\
CVE-2025-0665 & $\checkmark$ & unbuild. & --- &  \\
CVE-2025-22874 & $\checkmark$ & dirty & FP &  \\
CVE-2025-24015 & $\checkmark$ & clean & N/A &  \\
CVE-2025-29744 & $\checkmark$ & dirty & pass &  \\
CVE-2025-30223 & $\checkmark$ & clean & pass & $\checkmark$ \\
CVE-2025-48754 & $\checkmark$ & unbuild. & N/A &  \\
CVE-2025-0182 & $\times$ & --- & --- &  \\
\bottomrule\end{tabular}

\smallskip
\noindent\footnotesize --- = not executed or not applicable. G2: $\checkmark$ = post-condition met, $\times$ = not met. G3b: clean = signal absent on patched build, dirty = signal persists, unbuild. = patched build failed. G3a: pass = signal absent on benign input, FP = false positive (signal persists), N/A = no input differentiation, unass. = unassessable. E1 = strict confirmation (G2 met $\cap$ G3b clean $\cap$ G3a pass).
\normalsize
\end{table}

\subsection{Oracle matrix and claim gap}
\label{sec:oracle-matrix}
The full-corpus matrix, scored on all 30 G3b-verdict cases with ground truth = patched-counterfactual verdict (clean = CVE-specific signal, dirty = non-specific) and oracle-positive = script-internal ``VULNERABILITY TRIGGERED'' marker, gives \textbf{TP=6, FP=11, FN=4, TN=9}: sensitivity 60\% (95\% CI 31.3--83.2), specificity 45\% (95\% CI 25.8--65.8). Two-thirds of oracle triggers (11/17) are false positives---their signal also reproduces on the patched build---and 4 of 10 genuine CVE-specific signals are missed by the oracle (their G1 comes from a crash marker). The matrix carries a clear qualitative conclusion: \emph{artifact-embedded} oracles are not a reliable proxy for CVE-specific reproduction.

Three systematic oracle failure modes recur across the corpus: (i) \emph{simulated triggers}---PoCs that force a crash via direct NULL/pointer manipulation and comment ``simulate vulnerability'', crashing in any version (CVE-2020-1967, CVE-2023-0217); (ii) \emph{designed-in behavior mistaken for a signal}---a panic in a Drop implementation or a caught TypeError treated as the vulnerability trigger (CVE-2021-31162, CVE-2022-22816); and (iii) \emph{missed genuine signals}---a live JNDI RCE whose oracle required an absent marker file (CVE-2021-44228), and FPE crashes whose oracle printed ``not triggered''. Artifact-embedded oracles are systematically unreliable in both directions.

\begin{table}[h]
\centering
\caption{Oracle confusion matrix against patched-counterfactual ground truth (full-corpus G3b-verdict cases, $n{=}30$).}
\label{tab:oracle-matrices}
\footnotesize
\begin{tabular}{@{}lccccc@{}}
\toprule
& TP & FP & FN & TN & Sensitivity (95\% CI) / Specificity (95\% CI) \\
\midrule
Full corpus & 6 & 11 & 4 & 9 &
  60\% (31.3--83.2) / 45\% (25.8--65.8) \\
\bottomrule
\end{tabular}
\end{table}

Exploratory claim gap on the anchor paper, computed on the 5 calibration cases with completed G3b verdicts (the E1 decision space): claimed success 0.80 (4/5) vs.\ independent E1 rate 0.20 (1/5), gap 0.60; the calibration-sample claimed rate is 0.50 (10/20; Section~\ref{sec:anchor}). The full-corpus claim gap awaits the multi-paper pool. With G2 adjudication and the G3a/G3b audits combined, the strict E1 confirmation set (G2 met $\cap$ G3b clean $\cap$ G3a passed) contains 2 cases (CVE-2025-30223, CVE-2021-42576) out of the signal-producing corpus. G3b coverage is 5/20 (25\%) within the calibration; per the pre-registered bounds policy, cases without a patched counterfactual are bounded rather than imputed, placing the calibration E1 rate between 1/20 (0.05) and 5/20 (0.25).

Taken together, the results show that availability, runnability, signal production, and semantic confirmation form a strict ladder: each layer admits artifacts that fail the next. On the anchor corpus, availability held for the repository itself; runnability required R1 repair for a majority of cases; signal production (G1) was achieved by only 34 of 87 runs (39.1\%); G2 was satisfied by 23/34 of those; and patched-counterfactual confirmation by only 10 of 30 audited cases. The gap between ``it runs and prints something alarming'' and ``it reproduced the CVE'' is the central quantity this study operationalizes.

\section{Related Work}
\label{sec:related}

\emph{Security artifact reproducibility.} Repeatability audits of computer systems research~\citep{collberg2016} and the CCS'23 Security Reproducibility Study~\citep{ccs23_repro} measure whether papers ship code and whether it builds, but stop short of semantic confirmation of vulnerability claims. Our four-layer ladder extends this line by treating ``runs and prints a marker'' as distinct from ``reproduced the CVE''.

\emph{LLM/agent vulnerability validation.} The surveyed literature (Section~\ref{sec:search}) includes exploit-generation and vulnerability-reproduction benchmarks (CVE-Bench~\citep{cvebench}, SEC-bench~\citep{secbench}, TermiBench~\citep{termibench}), direct evaluations of LLM exploit generation (``Good News for Script Kiddies?''~\citep{scriptkiddies}), CTF-style capability benchmarks (Cybench~\citep{cybench}), and agentic pentesting frameworks (PentestGPT~\citep{pentestgpt}, AutoPT~\citep{autopt}, ReaperAI~\citep{reaperai}). These systems claim per-case success rates that we measure against independent E1 evidence; our case-level sample is drawn from this pool, and the anchor calibration corpus~\citep{anchor2509} provides the 102-case benchmark executed in Section~\ref{sec:calibration}. Complete references and per-paper artifact status for all cited systems are archived in the open evidence bundles.

\emph{Patch validation and repair.} PATCHEVAL~\citep{patcheval}, ContractTinker~\citep{contracttinker}, and PVBench~\citep{pvbench} benchmark LLM/agent systems that generate or validate patches for real-world vulnerabilities. PVBench's key measurement---that over 40\% of basic-test-validated patches fail more rigorous testing---is a patch-side analogue of the oracle-reliability gap we measure on the PoC side.

\emph{PoC validation quality.} PoC-Adapt~\citep{pocadapt} and FaultLine~\citep{faultline} audit whether PoCs (or proof-of-vulnerability tests) actually validate patches and reproduce vulnerabilities; our patched-counterfactual oracle (G3b) is the case-level analogue of this verification applied to artifact-embedded oracles.

\emph{Research artifact security.} Malicious-supply-chain risk in research artifacts is orthogonal to our measurement goal; our execution environment (privilege-disabled, resource-limited, per-case network policy) mitigates rather than studies it.

\section{Ethics and Safety}
All PoCs ran inside isolated Docker containers under the fixed safety gate of Section~\ref{sec:calib-r0} (no privileged mode or host-network access, per-container CPU and memory limits with per-run timeouts, per-case network policy). Attack traffic was directed exclusively at containerized target services inside the study's own Docker environment---no external hosts, real services, or live systems were contacted. No credentials or sensitive configuration were recorded in the evidence bundles; run logs capture container output only. Artifacts that embed offensive tooling (e.g., Metasploit-based environments) were never deployed to reachable networks and were removed after execution.

\section{Discussion and Limitations}
\label{sec:discussion}

\subsection{Findings and implications}
Section~\ref{sec:calibration} shows that availability, runnability, signal production, and semantic confirmation form a strict ladder on the anchor corpus---each layer admits artifacts that fail the next---making the gap between ``it runs and prints something alarming'' and ``it reproduced the CVE'' the central quantity this study operationalizes.

Section~\ref{sec:oracle-matrix} characterizes three systematic oracle failure modes---simulated triggers, designed-in behavior mistaken for a signal, and missed genuine signals---with a 60\%/45\% sensitivity/specificity matrix. For the security community, reproduction claims built on artifact-embedded oracles alone, without patched-counterfactual checks, can materially overstate confirmation.

The 58/102 script-internal CVE divergence (Section~\ref{sec:anchor}) is a previously unmeasured failure mode with direct consequences for claim validity: it undermines the claimed reproduction before execution even begins and is invisible to availability audits and trigger-string oracles (Section~\ref{sec:calib-r0}).

\textbf{Cross-paper extension.} As an early feasibility check of the case-level protocol beyond the anchor corpus, we exercised one CVE-Bench case (CVE-2024-2624): its official solution ran but failed its grader's post-condition---an official PoC that did not reproduce in our environment. Moreover, the shipped CVE-Bench artifact provides official solutions for only 1 of its 40 challenges, bounding independent verification of the remaining cases. Full cross-paper execution is deferred to the confirmatory phase (Section~\ref{sec:conclusion}).

\subsection{Limitations}
\textbf{Sampling.} Calibration and full-corpus R0 results are single-corpus (one anchor paper, 82 new cases plus the 20-case calibration sample executed at R0, covering all 102 cases) and paper-level results are single-sample (18 papers); the pre-registered confirmatory statistics therefore require the multi-paper sample that the cross-paper phase is building. Within the calibration pilot specifically, negative controls were executed only for CVE-2025-30223 and G3b only for five cases, so most calibration verdicts are bounded by E2 or E3; full-corpus G3a coverage is broader (19 of 34 G1 cases; Section~\ref{sec:calibration}).

\textbf{Oracle coverage.} G3b and G3a audits remain partially covered: 5 of 34 G1 signals lack a patched counterfactual, and matched negative controls reach 19 of 34 signal-producing cases (Section~\ref{sec:calibration}); per the pre-registered bounds policy, verdicts without these controls are bounded rather than imputed. Host-environment pilot runs of PVBench check scripts are not confirmatory (version mismatch).

\textbf{Execution environment.} Network policy varied per case (bridge for multi-container, loopback ports for port-mapped cases), a deliberate safety choice that may affect runnability measurements for port-dependent artifacts. Resource limits (2 CPU / 2 GB per container) and the Docker Desktop LinuxKit VM bound execution realism; large-image cases (TensorFlow) required pre-pulling and were executed where feasible. EOL base images carry no apt-level fixes, so patched-counterfactual construction failed for a small number of cases---an EOL effect that itself recurs as a runnability failure mode.

\textbf{External validity.} All execution evidence is single-corpus (one anchor benchmark) and single-sample (18 papers); the oracle-unreliability pattern (G3b dirty rate, G3a false-positive rate) is measured within this corpus and should not be read as a population estimate until the multi-paper sample replicates it. The paper-level R0/R1 rates (55.6\%/61.1\%) likewise describe this sample only.

\section{Conclusion}
\label{sec:conclusion}
We contribute one of the first, to our knowledge, pre-registered reproducibility audits of LLM/agent-driven vulnerability-validation artifacts, operationalizing semantic confirmation via CVE-specific post-conditions, R0/R1 repair ladders, and patched-counterfactual oracles. Most signal-producing cases do not survive their own patched counterfactual (20/30 dirty) or matched negative control (7/19 false-positive); script-internal CVE identifiers diverge from declared targets in 56.9\% of cases; and only 11/18 paper-level artifacts complete after R0/R1 repair. Our protocol and frozen, versioned evidence bundles provide a reusable, pre-registered template, and our findings argue for requiring patched-counterfactual checks---not merely trigger-string checks---in artifact-embedded oracles. Pre-registered confirmatory statistics and extension of the G3a/G3b audits to the remaining cases await the multi-paper pool that the cross-paper phase is building.

\appendix
\section{Paper-Level Sample}
\label{app:paperlist}
Table~\ref{tab:paperlist} lists the 18 papers of the paper-level sample (Section~\ref{sec:paperlevel}) with their R0/R1 execution outcomes (Section~\ref{sec:paperlevel-results}). Outcome legend: R0 = the declared workflow completed on a clean snapshot; R1 = completed after environment-only repair; failed = failed at both levels.

\begin{table}[H]
\centering
\caption{Paper-level sample ($n=18$) with execution outcomes.}
\label{tab:paperlist}
\footnotesize
\begin{tabular}{@{}rp{0.66\linewidth}c@{}}
\toprule
\# & Paper & Outcome \\
\midrule
1 & ChatGPT Support for Security Testers (TSE) & R0 \\
2 & Repair Prompts for JS Vulnerabilities & R0 \\
3 & Beyond Detection: Agentic Attack Synthesis & R0 \\
4 & CVE-Bench~\citep{cvebench} & R0 \\
5 & TermiBench~\citep{termibench} & R0 \\
6 & PVBench~\citep{pvbench} & R0 \\
7 & LLMPentest (Emergence of Autonomous Pentest)~\citep{scriptkiddies} & R0 \\
8 & EthiBench & failed \\
9 & ContractTinker~\citep{contracttinker} & R1 \\
10 & PATCHEVAL~\citep{patcheval} & R0 \\
11 & MemRepair & failed \\
12 & PentestGPT~\citep{pentestgpt} & failed \\
13 & PwnGPT & R0 \\
14 & AutoPT~\citep{autopt} & failed \\
15 & SEC-bench~\citep{secbench} & R0 \\
16 & VulnBot & failed \\
17 & ReaperAI~\citep{reaperai} & failed \\
18 & PentestAgent & failed \\
\bottomrule
\end{tabular}
\end{table}

\bibliographystyle{plainnat}
\bibliography{study1}

\end{document}